\documentclass[a4paper,twoside]{article}
\usepackage{etex}
\usepackage[T1]{fontenc}
\usepackage[latin1]{inputenc}
\usepackage[english]{babel}
\usepackage{graphicx}
\usepackage[noindentafter,pagestyles]{titlesec}
\usepackage{textcomp}
\usepackage{dcolumn}
\usepackage{amsmath}
\usepackage{txfonts}
\usepackage{fancyhdr}
\usepackage{threeparttable}
\usepackage{xspace}
\usepackage[superscript]{cite}
\usepackage{pictex}
\newcommand{\bmp}[1]{\begin{minipage}{#1\columnwidth}}
\newcommand{\emp}{\end{minipage}}

\newcommand{\bea}{\begin{eqnarray}}
\newcommand{\eea}{\end{eqnarray}}
\newcommand{\be}{\begin{equation}}
\newcommand{\ee}{\end{equation}}

\usepackage{duckuments}
\usepackage{tikz}
\usetikzlibrary{arrows.meta, positioning, automata, calc, shapes.geometric}
\usepackage{algorithm}
\usepackage{algpseudocode}
\usepackage{subcaption}
\usepackage{multirow}
\usepackage{booktabs}

\titleformat{\section}{\large\bfseries}{\thesection.}{.3em}{}
\titlespacing*{\section}{\leftmargini}{*3}{*3}
\titleformat{\subsection}{\bfseries}{\thesubsection}{.3em}{}
\titlespacing*{\subsection}{0pt}{*3}{*3}
\makeatletter
\def\@maketitle{%
  \newpage
  \null
  \vskip 2em%
  \begin{center}%
  \let \footnote \thanks
    {\fontsize{18}{22}\fontseries{b}\selectfont \@title \par}%
    \vskip 1.5em%
    {\normalsize
      \lineskip .5em%
      \begin{tabular}[t]{c}%
\@author
      \end{tabular}\par}%
    \vskip 1em%
    {\large \@date}%
  \end{center}%
  \par
  \vskip 1.5em}
\renewenvironment{abstract}{%
\if@twocolumn
\section*{\abstractname}%
\else
\quotation
\noindent{\bfseries\large \abstractname\vspace*{.3ex}\par}
\fi}
{\if@twocolumn\else\endquotation\fi}
\makeatother

\fancypagestyle{plain}{%
\fancyhf{} 
\fancyhead[L]{\small \MakeUppercase{\hspace*{15mm} 11$^{\mbox{\tiny th}}$ European Conference for Aeronautics and AeroSpace Sciences (EUCASS)} \\ \quad \\ DOI: ADD DOINUMBER HERE}
\fancyfoot[L]{\small Copyright \copyright\ 2025 by U.Zucchelli, M. A. Mendez, A. Urbano, S. V. Bonnieu, P.Wenderski, F. Sanfedino. Published by the EUCASS association with permission.}%
} 
\begin{document}
%
%
%
%
%
\title{Closed-loop control of sloshing fuel in a spinning spacecraft}
\author{\itshape U. Zucchelli $^{\star\dag}$, M. A. Mendez $^{\star\star}$, A. Urbano $^\star$, S. V. Bonnieu $^{\star\star\star}$, P. Wenderski $^{\star\star\star}$, F. Sanfedino$^\star$\\
\itshape$^\star$ F\'ed\'eration ENAC ISAE-SUPAERO ONERA, Universit\'e de Toulouse, 10 Av. Marc P\'elegrin, 31055, Toulouse, France \\
$^{\star\star}$ von Karman Institute for Fluid Dynamics,
Environmental and Applied Fluid Dynamics Department \\
$^{\star\star\star}$ European Space Agency, ESTEC, Noordwijk, the Netherlands \\
umberto.zucchelli2@isae-supaero.fr\\
$^\dag$Corresponding author}

\date{}
\maketitle
\begin{abstract}
  \noindent 
New-generation space missions require satellites to carry substantial amounts of liquid propellant, making it essential to analyse the coupled control-structure-propellant dynamics in detail. While Computational Fluid Dynamics (CFD) offers high-fidelity predictions, its computational cost limits its use in iterative design. Equivalent Mechanical Models (EMMs) provide a faster alternative, though their predictive performance--especially in closed-loop scenarios--remains largely unexplored.
This work presents a comparative analysis of a spacecraft under feedback control, using both CFD and a reduced-order sloshing model. Results show good agreement, validating the simplified model for the manoeuvrer considered.
This validation enables efficient sensitivity and stability studies, offering a practical tool for early-stage spacecraft design.\end{abstract}

\section{Introduction}
The prediction and control of liquid sloshing in spacecraft have become key topics in recent years. Propellant motion can significantly affect the centre of mass and attitude dynamics, particularly in next-generation systems carrying large fuel loads, deploying flexible appendages, and operating in long-duration missions. In such cases, the non-linear coupling between rigid-body motion, sloshing in microgravity, structural vibrations, and actuator forces cannot be neglected \cite{liu_2018, simonini_2024, rodrigues_2025}. Managing this coupling is crucial during control system design to ensure structural integrity and attitude performance.

While high-fidelity Computational Fluid Dynamics (CFD) simulations can provide accurate force predictions, their computational cost often limits their use in control design or system-level studies. To address this, researchers rely on reduced-order Equivalent Mechanical Models (EMMs), or mechanical analog models, which approximate sloshing dynamics using simplified systems with few degrees of freedom. These include spring-mass systems \cite{Abramson1966,Dodge2000,Enright1994,Jang2013}, pendulum models \cite{Gasbarri2016,Yue2011,Kang2005,Kang2008}, constraint surface models \cite{Zhou2015, Elke2024}, and pulsating ball models \cite{Vreeburg1997,Deng2017}. Each model has specific strengths and limitations, with the constraint surface model showing promising agreement for large-amplitude sloshing in low-g environments.
EMMs are typically calibrated using open-loop high-fidelity data and then applied to closed-loop performance evaluations. However, their reliability under feedback control remains largely untested, raising concerns about their fidelity in such conditions.

This study addresses this gap by introducing a simulation framework to analyse the coupled sloshing-structure-control dynamics of a spacecraft. We model a rigid spacecraft with sloshing propellant under angular velocity feedback control, using an EMM based on a particle constrained to a revolute surface \cite{Zhou2015,Elke2024}. Model predictions are compared with high-fidelity CFD simulations performed using the DIVA software, previously validated against microgravity experiments \cite{Dalmon2019}. The comparison evaluates whether the EMM can replicate the sloshing-induced dynamics observed in the CFD results under closed-loop control. The following sections present the simulation tools, modelling approach, and the integrated framework used to assess control-structure-sloshing interactions.

\section{Methodology}

\subsection{CFD Numerical Solver}

The isothermal sloshing simulations in this study were carried out using the Direct Numerical Simulation solver \textsc{DIVA} (Dynamics of Interface for Vaporisation and Atomization). The code was originally developed for high-fidelity two-phase flow simulations and has been extensively validated for isothermal configurations~\cite{Tanguy2005, Lalanne2015}, as well as for two-phase flows with phase change, including droplet impact in the Leidenfrost regime~\cite{RuedaVillegas2017} and nucleate boiling~\cite{Huber2017, Urbano2018}. In particular, Dalmon et al.~\cite{Dalmon2019} validated \textsc{DIVA} against experimental measurements of isothermal sloshing in microgravity conditions, using data from the \textsc{FLUIDICS} experiment on the International Space Station~\cite{Mignot2017}.

In the present work, \textsc{DIVA} is used to simulate the fluid dynamics of a spacecraft tank undergoing prescribed manoeuvrers. The computational domain includes both fluid and solid regions, with the fluid domain further divided into liquid and gas subdomains. The motion of the liquid-gas interface is tracked using the Level-Set method~\cite{Osher1988, Sussman1994}, where the interface is defined as the zero level of a signed distance function~$\phi$. A separate level-set function~$\phi_s$ is used to represent the static solid boundary. The evolution of the interface is governed by the standard advection equation:

\begin{equation}
    \frac{\partial \phi}{\partial t} + \mathbf{u} \cdot \nabla \phi = \mathbf{0}
\end{equation}

\noindent where $\mathbf{u}$ is the fluid velocity field. To maintain the distance property of $\phi$ throughout the simulation, this transport equation is coupled with the reinitialization algorithm of Sussman et al. \cite{Sussman1994}.

\begin{figure}[!h]
\centering

\begin{minipage}[t]{0.48\textwidth}
    \centering
    \includegraphics[width=0.9\textwidth]{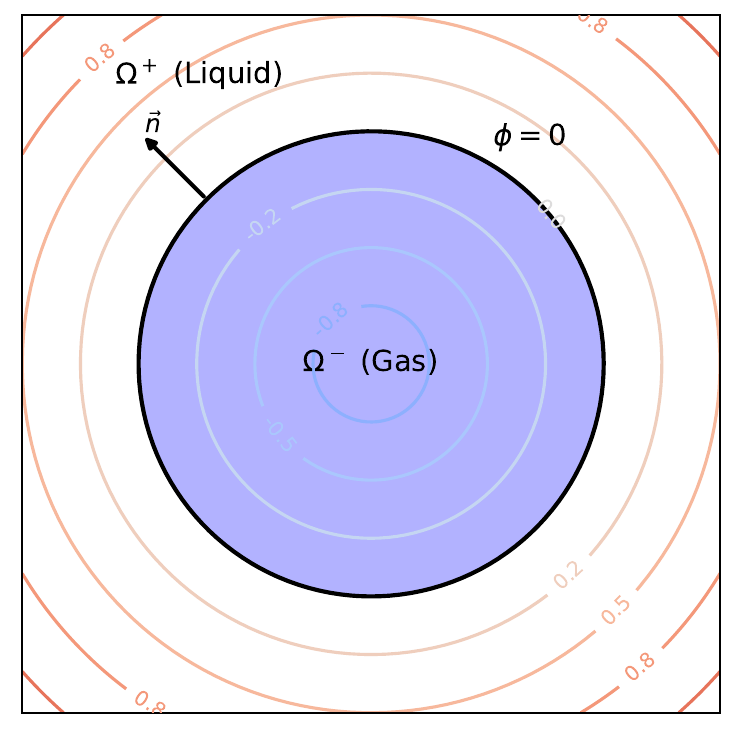}
    \caption{Level Set Function for a circular liquid-gas interface.}
    \label{Fig:level-set}
\end{minipage}
\hfill
\begin{minipage}[t]{0.48\textwidth}
    \centering
    \includegraphics[width=0.9\textwidth, trim=50 35 40 30, clip]{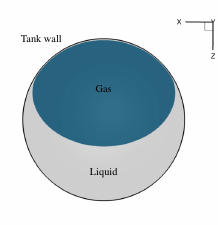}
    \caption{Solid-Gas-Liquid interface in a circular tank with 0-contact angle. Snapshot shows 3M NOVEC 2704 (Liquid) and air (Gas) under constant acceleration $g_z=-10^{-2}\,\mathrm{m/s}^2$, computed using DIVA.}
    \label{Fig:0contact-angle}
\end{minipage}

\end{figure}

To account for the three-phase contact line (liquid-solid-gas), additional boundary conditions are required for the scalar function $\phi$~\cite{Fedkiw1999, Aslam2004}. Following the approach adopted in~\cite{Dalmon2019}, this study imposes a zero contact angle condition to represent a perfectly wetting fluid. This is equivalent to assuming that a thin liquid layer remains in contact with the tank walls at all times (see Fig.~\ref{Fig:0contact-angle}).

The incompressible Navier-Stokes equations for two-phase flows are solved using the Ghost Fluid Conservative Viscous Method with an implicit time integration scheme (GFCMI)~\cite{Lepilliez2016}. Given the velocity field $\mathbf{u}^n$ at time $t = n \Delta t$, the updated field $\mathbf{u}^{n+1}$ is obtained through a three-step procedure.

First, an intermediate velocity field $\mathbf{u}^*$ is computed at $t^{n+1}$ by solving the momentum equation without the pressure gradient term, but including volume forces:

\begin{equation}
\rho\left(\frac{\mathbf{u}^* - \mathbf{u}^n}{\Delta t} + (\mathbf{u}^n \cdot \nabla)\mathbf{u}^n - \frac{\mathbf{F}^{vol}}{\rho}\right) = \nabla \cdot (2\mu\overline{\overline{\mathbf{D}}}^{*})\,,
\label{eq:momentum-eq}
\end{equation} where $\rho$ and $\mu$ are the density and viscosity, $\mathbf{F}^{vol}$ represents external volume forces, and $\overline{\overline{\mathbf{D}}}^{*}$ is the rate-of-deformation tensor for $\mathbf{u}^*$, treated implicitly to eliminate the timestep restriction from viscous terms.

Second, the pressure field $p^{n+1}$ is obtained by solving a Poisson equation with Neumann boundary conditions:

\begin{equation}
\nabla \cdot \left(\frac{\nabla p^{n+1}}{\rho}\right) = \frac{\nabla \cdot \mathbf{u}^*}{\Delta t} + \nabla \cdot \left(\frac{\sigma\kappa\mathbf{n}\delta_{\Gamma}}{\rho}\right)\,,
\end{equation}

The surface tension contribution $\sigma \kappa \mathbf{n}\delta_\Gamma$ accounts for capillary forces, where $\sigma$ is the surface tension coefficient, $\kappa$ is the interface curvature, $\mathbf{n}$ the unit normal vector, and $\delta_\Gamma$ a Dirac distribution centered on the liquid-gas interface $\Gamma$. Interface sharpness and pressure jumps are incorporated using a sharp interface representation following Liu et al.~\cite{Liu2000}.

Third, the final velocity $\mathbf{u}^{n+1}$ is recovered by projecting the intermediate velocity onto a divergence-free field:

\begin{equation}
\mathbf{u}^{n+1} = \mathbf{u}^* - \frac{\Delta t}{\rho}(\nabla p^{n+1} - \sigma\kappa\mathbf{n}\delta_{\Gamma})\,.
\end{equation}

No-slip boundary conditions at solid walls are enforced using a subcell resolution method, primarily following Lepilliez et al.\cite{Lepilliez2016}, Gibou et al. \cite{Gibou2002}, and Ng et al. \cite{Ng2009}. The resulting sloshing force is computed as the volume integral of the pressure and surface tension forces acting on the fluid domain:

\begin{equation}
\mathbf{F}_{slosh} = \int_{\Omega_f} \left(\nabla p - \sigma \kappa \mathbf{n}\delta_\Gamma \right) \, dV\,.
\label{eq:Fslosh}
\end{equation}

\subsection{Sloshing Model}
\begin{figure}[!h]
\centering

\begin{minipage}[t]{0.48\textwidth}
    \centering
    \includegraphics[width=0.9\textwidth]{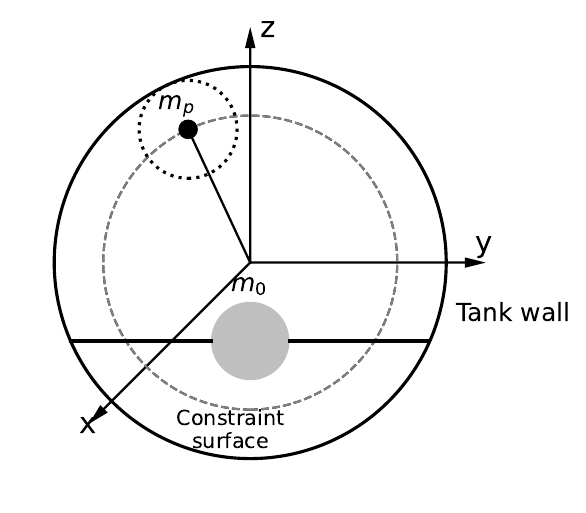}
    \caption{Schematic diagram of the Equivalent Mechanical Model.}
    \label{Fig:model-sketch}
\end{minipage}
\hfill
\begin{minipage}[t]{0.48\textwidth}
    \centering
    \includegraphics[width=0.9\textwidth]{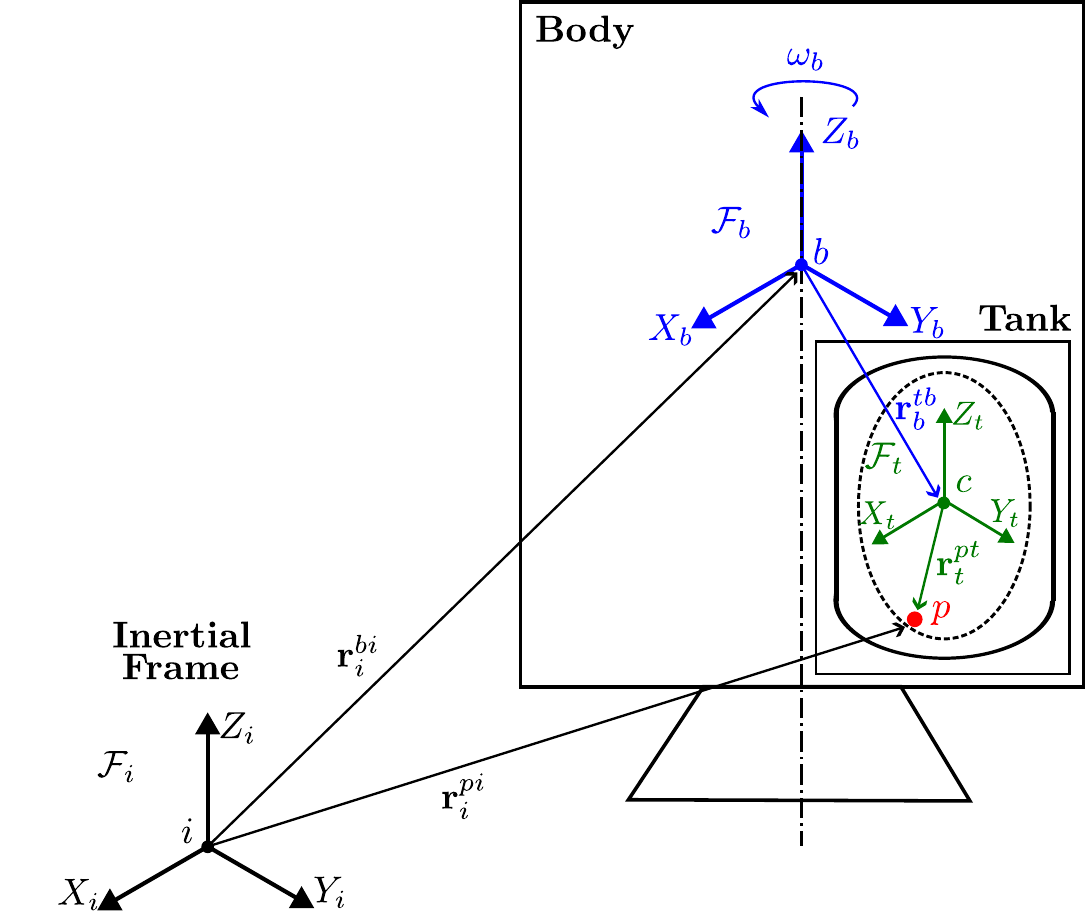}
    \caption{Sketch of reference frames and constraint surface model.}
    \label{Fig:ref-frames-model}
\end{minipage}

\end{figure}

We model liquid sloshing in microgravity using a mechanical analog based on the constraint surface method~\cite{Zhou2015, Elke2024}, which captures large-amplitude oscillations by representing the propellant as a single particle moving along a predefined geometry. Since only a fraction of the fluid contributes dynamically to the sloshing motion, the total propellant mass $ m_{l,\text{tot}} $ is partitioned into a stationary component $ m_0$ and a moving mass $ m_p$, which follows the constraint surface, as illustrated in Fig.~\ref{Fig:model-sketch}.

To describe the particle motion, three reference frames are introduced, as shown in Fig.~\ref{Fig:ref-frames-model}: the inertial frame \(\mathcal{F}_i\), the body-fixed frame \(\mathcal{F}_b\) located at the center of the spacecraft hub, and the tank-fixed frame \(\mathcal{F}_t\), which defines the geometry of the constraint surface. The position and velocity vectors are denoted by \(\mathbf{r}\) and \(\mathbf{v}\), respectively, with time derivatives indicated by \(\dot{(\cdot)}\). Subscripts \(_i\), \(_b\), and \(_t\) refer to the inertial, body, and tank frames, respectively. Superscripts \({}^{rs}\) denote quantities related to element \(s\) with respect to element \(r\), where \(i\), \(b\), \(c\), and \(p\) indicate the inertial frame, body frame origin, tank center, and moving particle. Frame transformations are performed using Direction Cosine Matrices (DCMs); for example, \(\mathbf{C}_{t \leftarrow b}\) transforms a vector from the body frame \(\mathcal{F}_b\) to the tank frame \(\mathcal{F}_t\).

Assuming the tank is axisymmetric, the constraint surface is defined as a solid of revolution by the equation:
\begin{equation}
    S_i : \frac{x^2}{a_i^2} + \frac{y^2 + z^2}{b_i^2} - 1 = 0\,.
\end{equation}

The model operates in two regimes: an unconstrained mode, where the particle moves freely within the tank, and a constrained mode, where it remains in contact with the constraint surface. In both cases, the dynamics of the sloshing mass \( m_p \) are governed by Newton's second law, with the mass matrix \(\mathbf{M} = m_p \mathbf{I}_3\), yielding the inertial-frame equation of motion:
\begin{equation}
\mathbf{M} \dot{\mathbf{v}}_i^{pi} = \mathbf{F}\,.
\label{eq:newton_inertial}
\end{equation}

The total external force $ \mathbf{F} $ acting on the particle consists of three main contributions: gravitational, friction and contact force. The gravitational force is given by $ \mathbf{f}_g = \mathbf{M} \mathbf{g} $, where $\mathbf{g} \in \mathbb{R}^3 $ is the gravitational acceleration vector in the inertial frame. The friction force is modelled as a viscous damping term acting tangentially to the constraint surface to capture energy dissipation due to internal fluid motion and viscous losses at the wall:
\begin{equation}
    \mathbf{f}_f = - \frac{C_f \mu m_p}{\left( R_t - \|\mathbf{r}_t^{pc}\| \right)^2} \mathbf{v}_{t,\mathbin{//}}^{pc}\,,
    \label{eq:friction-term}
\end{equation} where $ C_f \in \mathbb{R}^+ $ is a tunable friction coefficient calibrated from CFD, $ \mu$ is the liquid's dynamic viscosity, $ R_t$ is the tank radius, and $ \mathbf{v}_{t,\mathbin{//}}^{pc} $ is the tangential velocity component in the tank frame. Lastly, the constraint force $ \mathbf{f}_c \in \mathbb{R}^3 $ acts normal to the surface at the contact point and is introduced via a Lagrange multiplier to enforce the geometric constraint of surface confinement.

 In the unconstrained condition, contact and viscous forces $\mathbf{f}_f, \mathbf{f}_c$ vanish, and the particle is subject solely to gravity, resulting in the simplified dynamics:

\begin{equation}
    \mathbf{M} \mathbf{\dot v}_i^{pi} = \mathbf{F}=\mathbf{f_g}= \mathbf{M} \mathbf{g}\,.
\end{equation} 

In the constrained condition, the particle slides along the surface $S_i$, and its position \(\mathbf{r}_t^{pc}\) must satisfy the geometric constraint:

\begin{equation}
    \mathbf{C} \left( \mathbf{r}_t^{pc} \right) = {\mathbf{r}_t^{pc}}^T \mathbf{W}_t \mathbf{r}_t^{pc} - 1 = 0\,, 
\end{equation} with $\mathbf{W}_t = \mbox{diag}(a_i^{-2}, b_i^{-2}, b_i^{-2})$. Differentiating $\mathbf{C}(\cdot)$ yields the velocity-level constraint:  

\begin{equation}
\label{eq_C_dot}
    \dot{\mathbf{C}} = 2 {\mathbf{r}_t^{pc}}^T\mathbf{W}_t \mathbf{\dot r}_t^{pc} = \boldsymbol{\Lambda} \boldsymbol{\nu} = 0\,,
\end{equation} with generalised velocity

\begin{equation}
\boldsymbol{\nu} =
\bigl[{\mathbf{v}_i^{bi}}^T\;\boldsymbol{\omega}_{b}^T\;
      {\mathbf{v}_i^{pi}}^T\bigr]^T\,,
\end{equation}

\noindent where $\boldsymbol{\omega}_{b}$ is the angular velocity of the spacecraft with respect to the inertial frame, expressed in the body frame. The Jacobian $\boldsymbol{\Lambda}$ in \eqref{eq_C_dot} reads:

\begin{equation}
\boldsymbol{\Lambda}=
2{\mathbf{r}_t^{pc}}^T \mathbf{W}_t
\begin{bmatrix}
-\mathbf{C}_{t\leftarrow i} &
(\mathbf{r}_t^{pc})^\times\mathbf{C}_{t\leftarrow b}
            +\mathbf{C}_{t\leftarrow b}(\mathbf{r}_b^{cb})^\times &
\mathbf{C}_{t\leftarrow i}
\end{bmatrix}.
\label{eq:Lambda}
\end{equation}

In this framework, $\dot{\mathbf{v}}_i^{bi}$ and $\dot{\boldsymbol{\omega}}_{b}$ are considered known inputs of the model,
 and the only unknown is $\dot{\mathbf{v}}_i^{pi}$. Thus, we can partition:
\begin{equation}
    \boldsymbol{\nu} = 
    \begin{bmatrix}
    \boldsymbol{\nu}_k \\
    \mathbf{v}_i^{pi}
    \end{bmatrix},
    \quad
    \boldsymbol{\Lambda} = 
    \begin{bmatrix}
    \boldsymbol{\Lambda}_k & \boldsymbol{\Lambda}_p
    \end{bmatrix}\,.
    \label{eq:nu-Lambda}
\end{equation}

 Differentiating again the velocity constraint gives the acceleration-level form:
\begin{equation}
\ddot{C} = \boldsymbol{\Lambda} \dot{\boldsymbol{\nu}} + \dot{\boldsymbol{\Lambda}} \boldsymbol{\nu} = \boldsymbol{\Lambda}_p \mathbf{\dot v}_i^{pi} + \underbrace{\boldsymbol{\Lambda}_k \dot{\boldsymbol{\nu}}_k + \dot{\boldsymbol{\Lambda}}_p \mathbf{v}_i^{pi} + \dot{\boldsymbol{\Lambda}}_k \boldsymbol{\nu}_k }_{\displaystyle \beta} = 0,
\label{eq:acc_constraint}
\end{equation} where the terms $\mathbf{\dot \Lambda}$ take into account for the non-inertial accelerations, i.e. the Coriolis, Euler and centrifugal accelerations. Substituting the Lagrange representation
$\mathbf{f}_c=\boldsymbol{\Lambda}_p^\top\lambda$
into \eqref{eq:newton_inertial} gives
\begin{equation}
\mathbf{M} \mathbf{\dot v}_i^{pi} 
    = \mathbf{f}_g + \mathbf{f}_f + \boldsymbol{\Lambda}_p^\top\lambda\,.
\end{equation}
Combining this with the scalar acceleration constraint~\eqref{eq:acc_constraint} yields the augmented system:
\begin{equation}
\begin{bmatrix}
\mathbf{M} & -\boldsymbol{\Lambda}_p^\top \\[2pt]
-\boldsymbol{\Lambda}_p & 0
\end{bmatrix}
\begin{bmatrix}
\mathbf{\dot v}_i^{pi} \\ \lambda
\end{bmatrix}
=
\begin{bmatrix}
\mathbf{f}_g + \mathbf{f}_f \\[2pt] -\,\beta
\end{bmatrix}\,.
\label{eq:reduced_augmented}
\end{equation}

\noindent Since $\mathbf{M}$ is invertible, the $1\times1$ Schur complement yields a
closed-form solution for $\lambda$:

\begin{equation}
\lambda = -\,
             \frac{\boldsymbol{\Lambda}_p\mathbf{M}^{-1}(\mathbf{f}_g + \mathbf{f}_f)
                   \;+\; \beta}
                  {\boldsymbol{\Lambda}_p\mathbf{M}^{-1}\boldsymbol{\Lambda}_p^\top},
\qquad \text{and} \quad
\mathbf{\dot v}_i^{pi} = 
      \mathbf{M}^{-1}\!\bigl(\mathbf{f}_g + \mathbf{f}_f
      +\boldsymbol{\Lambda}_p^\top\lambda\bigr).
\end{equation}

 The simulation alternates between unconstrained and constrained modes, depending on contact conditions. The particle enters the constrained regime when the constraint equation is first satisfied, i.e., \( C(\mathbf{r}_t^{pc}) = 0 \) and the perpendicular velocity of the particle with respect to the surface is positive, i.e. \( \mathbf{v}_{t, \perp}^{pc} > 0 \). The event is modelled as a fully inelastic impact, so the resulting particle velocity is projected to match the surface normal velocity of the tank wall, while maintaining its original tangential velocity. 
 
The system exits the constrained regime via a separation event, triggered when the constraint force becomes tensile and exceeds a small adhesion threshold \( f_{\text{adh}} \in \mathbb{R}_+ \):
\begin{equation}
    -\mathbf{e}_n^T \mathbf{f}_c > f_{\text{adh}},
\end{equation}
\noindent where \( \mathbf{e}_n \) is the surface normal vector in the inertial frame.

\begin{figure}[!h]
\centering
\begin{tikzpicture}[
    node distance=3cm and 2cm,
    state/.style={
        rectangle,
        draw,
        text width=2cm,
        minimum height=1cm,
        text centered
    },
    sum/.style={
        inner sep=0pt
    },
    diamondstate/.style={
        draw,
        diamond,
        aspect=2,
        minimum height=0.5cm,
        text centered,
        text width=1.2cm
    },
    >=Latex
]
\node[state] (free) {Unconstrained};
\node[diamondstate] (impact) at (6,0) {Collision};
\node[state] (slide) at (12,0) {Constrained};
\node[sum] (sum_point) at (0,-2) {\scalebox{2}{$\otimes$}};
\node[diamondstate] (separation) at (6,-3) {Separation};

\coordinate (impact_bend) at ($(impact.south) + (0,-1.4)$);
\coordinate (sum_entry) at ($(sum_point) + (0.25, 0)$); 

\path[->] (free) edge node[above] {$C(\mathbf{r}_t^{pc}) = 0, \mathbf{v}_{t, \perp}^{pc} > 0$} (impact);
\path[->] (impact) edge node[above] {$C(\mathbf{r}_t^{pc}) = 0, \mathbf{v}_{t, \perp}^{pc} \geq 0$} (slide);

\draw[->] (free.south) -- ++(0, -0.5cm) -| node[near start, right] {$C(\mathbf{r}_t^{pc})>0$} (sum_point.north);
\draw[->] (impact.south) -- (impact_bend) -- (sum_entry)
    node[midway, above] {$C(\mathbf{r}_t^{pc}) = 0, \mathbf{v}_{t, \perp}^{pc} < 0$};
\draw[->] (slide.south) -- ++(0, -1.3cm) |- node[near start, left] {$-\mathbf{e}_n^T \mathbf{f}_c > f_{\text{adh}}$} (separation.east);
\draw[->] (separation.west) -| (sum_point.south);

\draw[->] (sum_point.west) -- ++(-2.5cm,0) |- (free.west);

\end{tikzpicture}
\caption{State diagram representing the working modes of the sloshing model, including transitions between unconstrained motion and constrained dynamics.}
\label{Fig:diagram-model}
\end{figure}
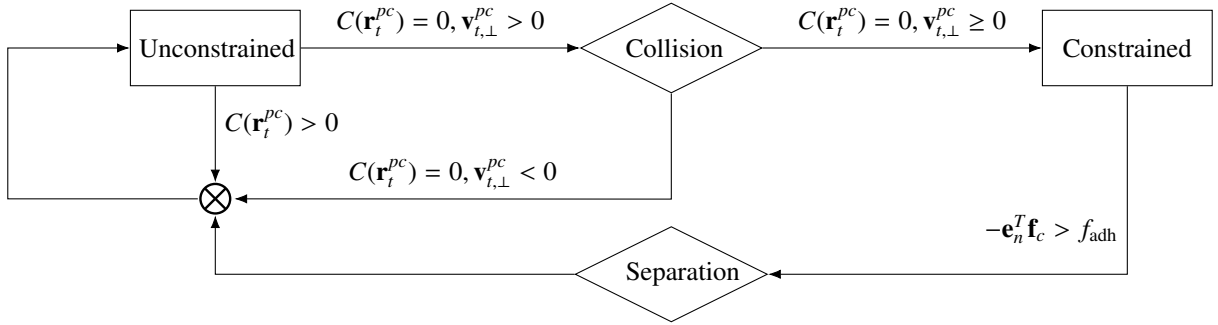

Figure~\ref{Fig:diagram-model} summarises the transition logic between these simulation modes.

\subsection{Coupling between satellite dynamics and sloshing model}

\begin{figure}[!h]
\centering
\resizebox{\textwidth}{!}{%
\begin{tikzpicture}[
    auto,
    node distance=1.8cm and 1.5cm,
    block/.style={
        rectangle,
        draw,
        fill=blue!20,
        text width=7em, 
        text centered,
        rounded corners,
        minimum height=1cm
    },
    sum/.style={
        circle,
        draw,
        fill=red!10,
        minimum size=0.7cm,
        node distance=1.8cm
    },
    line/.style={
        draw,
        -Latex
    }
]
\node[block] (guidance) {Guidance};
\node[sum, right=of guidance] (sum1) {};
\node[above left=0.05cm and 0.05cm of sum1] {\small $+$}; 
\node[below left=0.05cm and 0.05cm of sum1] {\small $-$}; 
\node[block, right=of sum1] (control) {Control};
\node[sum, right=of control] (sum2) {};
\node[above left=0.05cm and 0.05cm of sum2] {\small $+$}; 
\node[below left=0.05cm and 0.05cm of sum2] {\small $+$}; 
\node[block, right=of sum2] (satellite) {Satellite Dynamics};
\node[block, above=of satellite] (liquid) {Liquid};

\coordinate[right=1.5cm of satellite] (output_split);
\filldraw[black] (output_split) circle (3pt);
\path[line] (guidance) -- (sum1);
\path[line] (sum1) -- (control);
\path[line] (control) -- (sum2);
\path[line] (sum2) -- (satellite);
\path[draw] (satellite) -- (output_split);

\path[line] (output_split) -- ++(1.5,0) node[right] {System Output};
\path[line] (output_split) |- (liquid);
\path[line] (liquid) -| (sum2);
\path[line] (output_split) -- ++(0,-1.5) -| (sum1);
\end{tikzpicture}%
 }
\caption{System block diagram of the controlled closed-loop.}
\label{Fig:closed-loop-setup}
\end{figure}
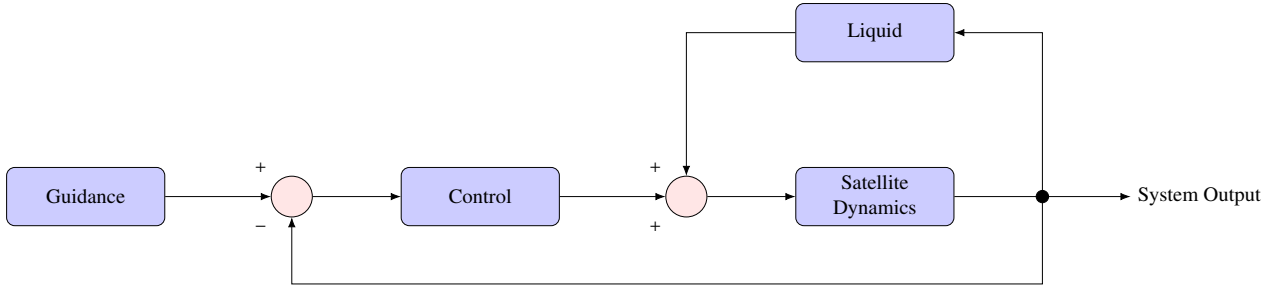

In this section, we present the developed simulation framework designed to study the interaction between propellant sloshing dynamics and spacecraft motion under closed-loop feedback control, as illustrated in Fig. \ref{Fig:closed-loop-setup}. 

\noindent The framework is intended to be general, supporting both control-structure-CFD coupling using DIVA and control-structure-mechanical analog coupling based on the constraint surface model.

In the CFD-based implementation, the standard algorithm in DIVA is adapted to compute the external volume force term $\mathbf{F}^{vol}$ (see Eq. \eqref{eq:momentum-eq}) based on provisional tank accelerations. These accelerations account for the control inputs applied at the current timestep and the sloshing disturbance forces from the previous step.  Using the estimated accelerations, DIVA solves the incompressible Navier-Stokes equations and updates the fluid motion, yielding the corresponding sloshing reaction forces through Eq.\eqref{eq:Fslosh}. These updated forces are then incorporated into the structural dynamics of the spacecraft, which are modelled as a fully nonlinear 3-DOF rotational system. Translational motion is neglected, as the focus of this study is on the coupling between sloshing and attitude dynamics. The rotational motion is integrated using quaternions for attitude tracking and the angular dynamics follow:
\begin{equation}
    \mathbf{J}_b \boldsymbol{\dot \omega}_b = \boldsymbol{\tau}_b - (\boldsymbol{\omega}_{b})^\times \mathbf{J}_b \boldsymbol{\omega}_{b}
\end{equation}
where $\boldsymbol{\tau}_{b}$ represents the external torques and includes both control torques and the sloshing-induced disturbances. Orbital perturbations are several orders of magnitude smaller than the sloshing-induced disturbances and are therefore neglected in this study. Finally, the updated rotational state is integrated to update the spacecraft dynamics and provide the necessary information for the controller at the next time step, closing the feedback loop. Algorithm 1, detailed in Figure \ref{Fig:simulation-algorithm}, summarises the principal steps of the proposed coupled simulations.

\begin{figure}[h!]
    \centering
    \begin{minipage}{0.95\linewidth}
        \begin{algorithm}[H]
        \caption{Coupled Fluid-Structure Sloshing Simulation}
        \begin{algorithmic}[1]
            \State \textbf{Initialize:} Set initial fluid and body states
            \State Define system properties (e.g., tank geometry, fluid parameters)
            \Procedure{Simulate}{$t_{\text{end}}, \Delta t$}
                \State $t \gets 0$
                \While{$t < t_{\text{end}}$}
                    \Comment{\textbf{Input}}
                    \State Retrieve sloshing state from previous timestep
                    \State Compute control torques from control law

                    \Comment{\textbf{Prediction Step}}
                    \State Compute provisional body acceleration
                    \State Solve fluid dynamics with provisional body motion
                    \State Evaluate updated sloshing forces and torques

                    \Comment{\textbf{Integration Step}}
                    \State Integrate rigid-body 6-DOF motion using current forces

                    \Comment{\textbf{Timestep Update}}
                    \State $t \gets t + \Delta t$
                \EndWhile
            \EndProcedure
            \State \textbf{Finalize:} Save simulation results for post-processing
        \end{algorithmic}
        \end{algorithm}
    \end{minipage}
    \caption{Algorithm outlining the coupled simulation of the controlled fluid-structure system with sloshing dynamics and 6-DOF rigid-body integration.}
    \label{Fig:simulation-algorithm}
\end{figure}

To ensure consistency and comparability, we adopt the same control-structure coupling algorithm for simulations using the mechanical analog model of sloshing. In this case, the estimated body accelerations required to evaluate sloshing forces at the current time step are introduced into the model as the vectors $\dot{\mathbf{v}}i^{bi}$ and $\dot{\boldsymbol{\omega}}{b}$, which appear in the generalized velocity vector $\boldsymbol{\nu}_k$ defined in Eq. \eqref{eq:nu-Lambda}. This unified framework enables closed-loop simulation of the coupled propellant-structure system using either high-fidelity DNS via DIVA or the mechanical analog model. As a result, it facilitates a direct comparison between the reduced-order model predictions and the detailed fluid-structure interaction results obtained from DIVA.

\section{Numerical results}
This section presents the validation and application of the proposed mechanical model for predicting sloshing forces. We first tune the model parameters and assess its performance under complex excitations that cause large liquid displacements. In Section~\ref{sec:open-loop-results}, we compare model predictions with results from the DIVA code in an open-loop setup, where the tank follows a prescribed acceleration profile. Section~\ref{sec:closed-loop-results} then examines the coupled structure--liquid system under closed-loop control, comparing sloshing forces from CFD simulations with those predicted by the mechanical model, within the previously introduced framework.

\subsection{Open-Loop Tuning and Validation}\label{sec:open-loop-results}

\begin{figure}[!h]
\centering
\begin{minipage}[t]{0.7\textwidth}
    \centering
    \includegraphics[height=90mm]{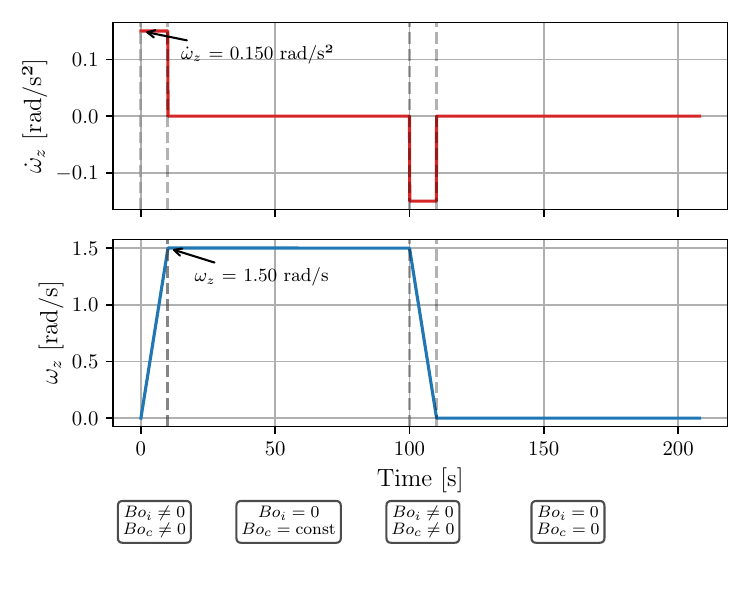}
    \caption{Angular acceleration and velocity profile of the manoeuvre.}
    \label{Fig:open-loop-manouver}
\end{minipage}
\hfill
\begin{minipage}[t]{0.25\textwidth}
    \centering
    \includegraphics[height=80mm]{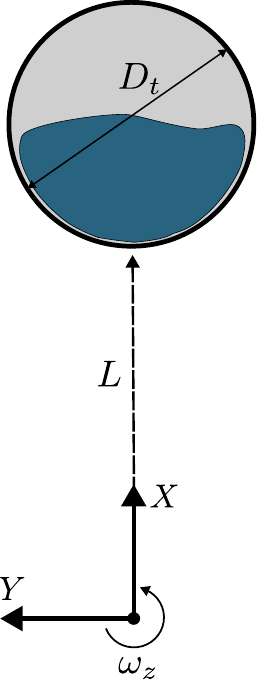}
    \caption{Schematic of the study case for the open-loop.}
    \label{Fig:skecth_OL}
\end{minipage}
\end{figure}

The DIVA code has been validated against microgravity data acquired on board the International Space Station during the FLUIDICS experiment \cite{Dalmon2019}. The experimental campaign investigated sloshing behaviour during flat-spin manoeuvres at varying angular velocities, acceleration profiles, and filling ratios.
To remain close the established DIVA validation range, which includes direct comparisons with experimental data, we select a flat-spin manoeuvre around the $z$ axis for the open-loop test case. The angular velocity profile is shown in Fig. \ref{Fig:open-loop-manouver}, alongside the schematic of the study case in Fig \ref{Fig:skecth_OL}. The maneuver begins with an acceleration phase $t_{acc}$, where angular velocity increases from zero to a maximum $\omega_b=\left[ 0, 0, \omega_z \right]$ with an angular acceleration $\dot \omega_b = \left[ 0, 0, \dot \omega_z \right]$. The tank then rotates at constant angular velocity to allow the fluid to stabilize, bofore returning to rest with a deceleration phase symmetric to the acceleration, with $t_{dec}=t_{acc}$ and $\dot \omega_{b,dec}= \dot \omega_b$.

The working fluid is 3M NOVEC 2704, the same fluid employed for code validation in Dalmon et al.\cite{Dalmon2019}, chosen for its physical properties that resemble propellants used in the space industry. Its fully wetting characteristic is modelled with a fixed liquid-wall contact angle of zero. 
In a low-gravity environment, two dimensionless numbers primarily govern the sloshing fluid dynamics: the Ohnesorge number (Oh) and the Bond number (Bo),

\begin{equation}
Oh = \sqrt{\frac{{\mu_l}^2}{\rho_l \sigma R_t}},
\qquad 
Bo = \frac{\rho_l \alpha {R_t}^2}{\sigma}.
\end{equation}

The Ohnesorge number quantifies the relative importance of viscous to surface tension effects, while the Bond number compares inertial forces arising from manoeuvre-induced acceleration to surface tension. It is useful to distinguish between two forms of the Bond number: the centripetal Bond number $Bo_c$, where $\alpha=\omega_z^2 L$, and the angular-acceleration Bond number $Bo_i$, where $\alpha=\dot \omega_z L$. $L=\left[ 0.2,0,0 \right]m$ is the distance of the centre of the tank from the rotation centre.
For this specific manoeuvre, fluid characteristics, and a spherical tank of diameter $D_t=0.1m$ filled to $50\%$ positioned at $L=0.2m$ from the spinning axis, the fluid dynamics is characterised by $Oh=1.021e^{-3}$, $Bo_c=116$, and $Bo_i=7.77$.

\begin{figure}[!h]
    \centering
    \begin{subfigure}[b]{0.75\textwidth}
        \centering
        \includegraphics[width=\textwidth]{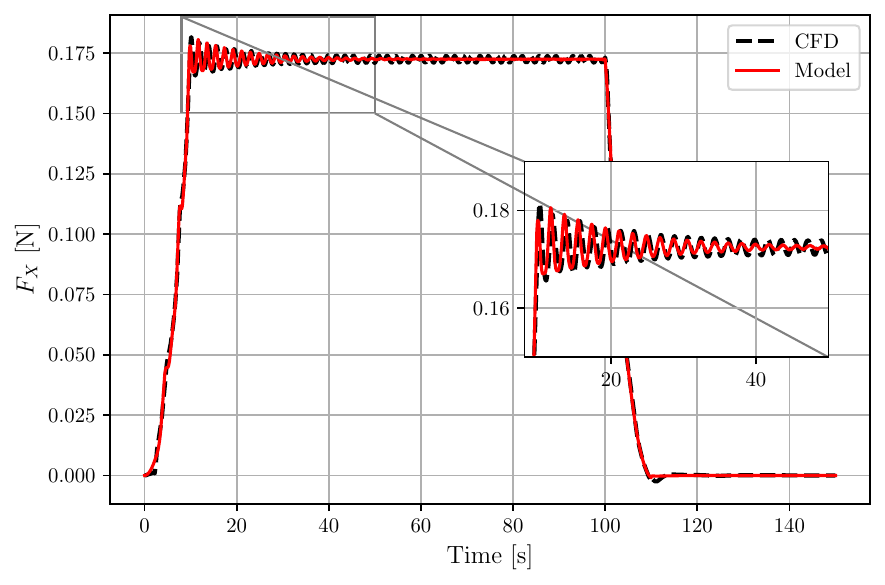}
        \caption{Force component $F_X$ from CFD and model.}
        \label{fig:sub1}
    \end{subfigure}
    \begin{subfigure}[b]{0.75\textwidth}
        \centering
        \includegraphics[width=\textwidth]{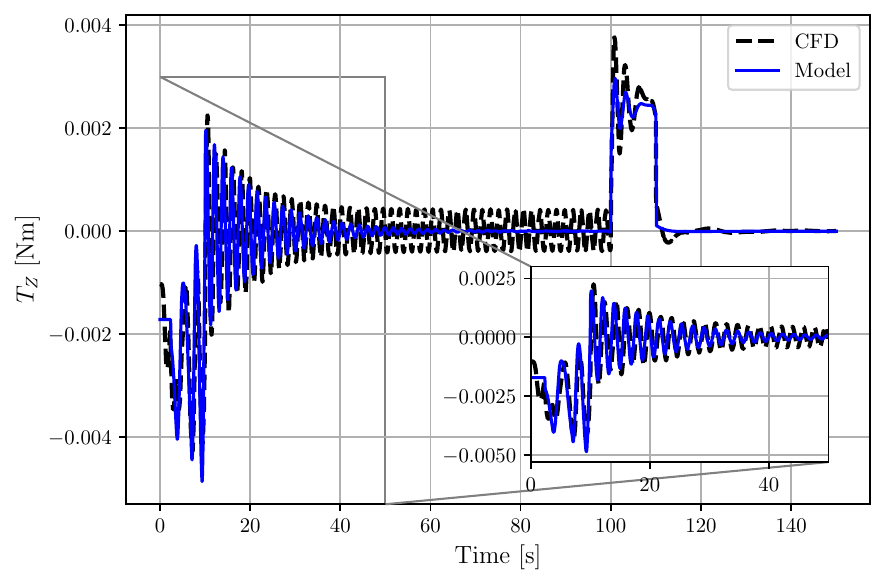}
        \caption{Torque component $T_Z$ from CFD and model.}
        \label{fig:sub3}
    \end{subfigure}
    \caption{Comparison of the main force and torque components from CFD and sloshing model during the open-loop manoeuvre.}
    \label{Fig:open-loop-results}
\end{figure}

For the EMM, three parameters are treated as unknowns: the characteristic radius of the constraint surface $a_i$ (with $a_i = b_i$ due to the spherical tank geometry), the friction coefficient $C_f$ in \eqref{eq:friction-term}, and the fraction of non-sloshing liquid mass $m_0$. The latter represents the portion of propellant assumed to remain fixed at the centre of the tank, while the remaining sloshing mass $m_p = m_{l,tot}-m_0$ is modelled using the constraint surface approach.
Their values are identified through a global optimisation process, using an open-source differential evolution algorithm from SciPy \cite{2020SciPy-NMeth}. The optimisation yields $m_0=78\%$ of the total propellant mass in the tank, $a_i=b_i=0.81 R_t$ and $C_f = 0.015$. Figure \ref{Fig:open-loop-results} presents the CFD results alongside the predictions from the tuned mechanical model.

\begin{figure}[!h]
    \centering
    \begin{minipage}{\textwidth}
        \centering
        \includegraphics[width=0.24\textwidth, trim=15 15 15 15, clip]{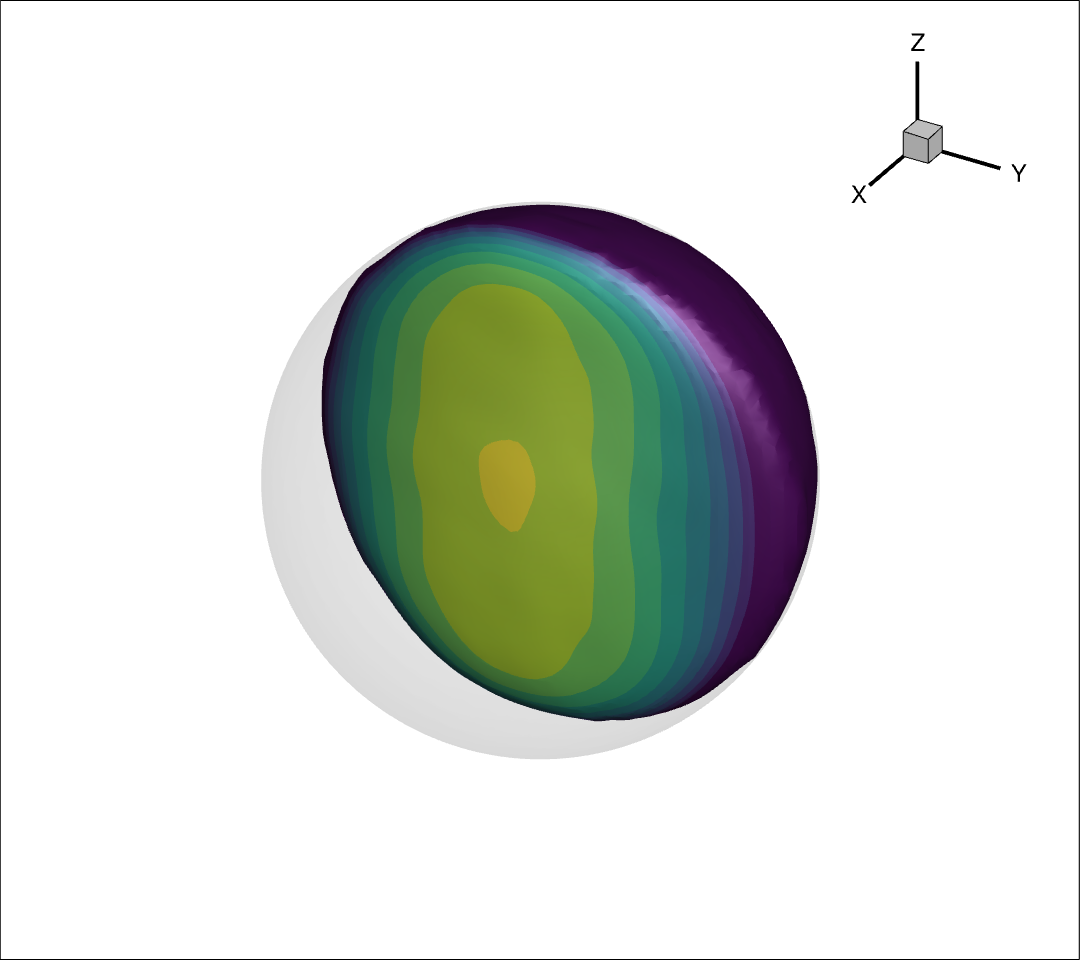}
        \includegraphics[width=0.24\textwidth, trim=15 15 15 15, clip]{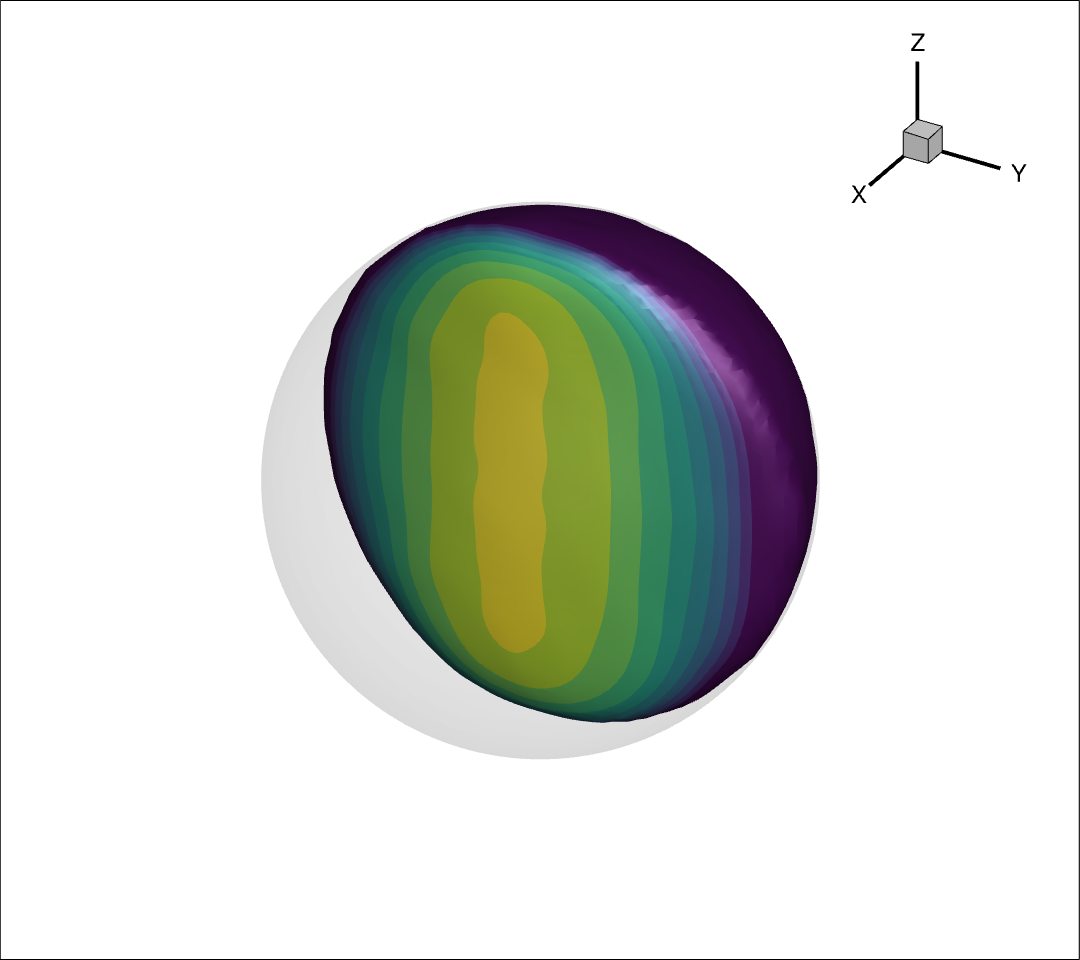}
        \includegraphics[width=0.24\textwidth, trim=15 15 15 15, clip]{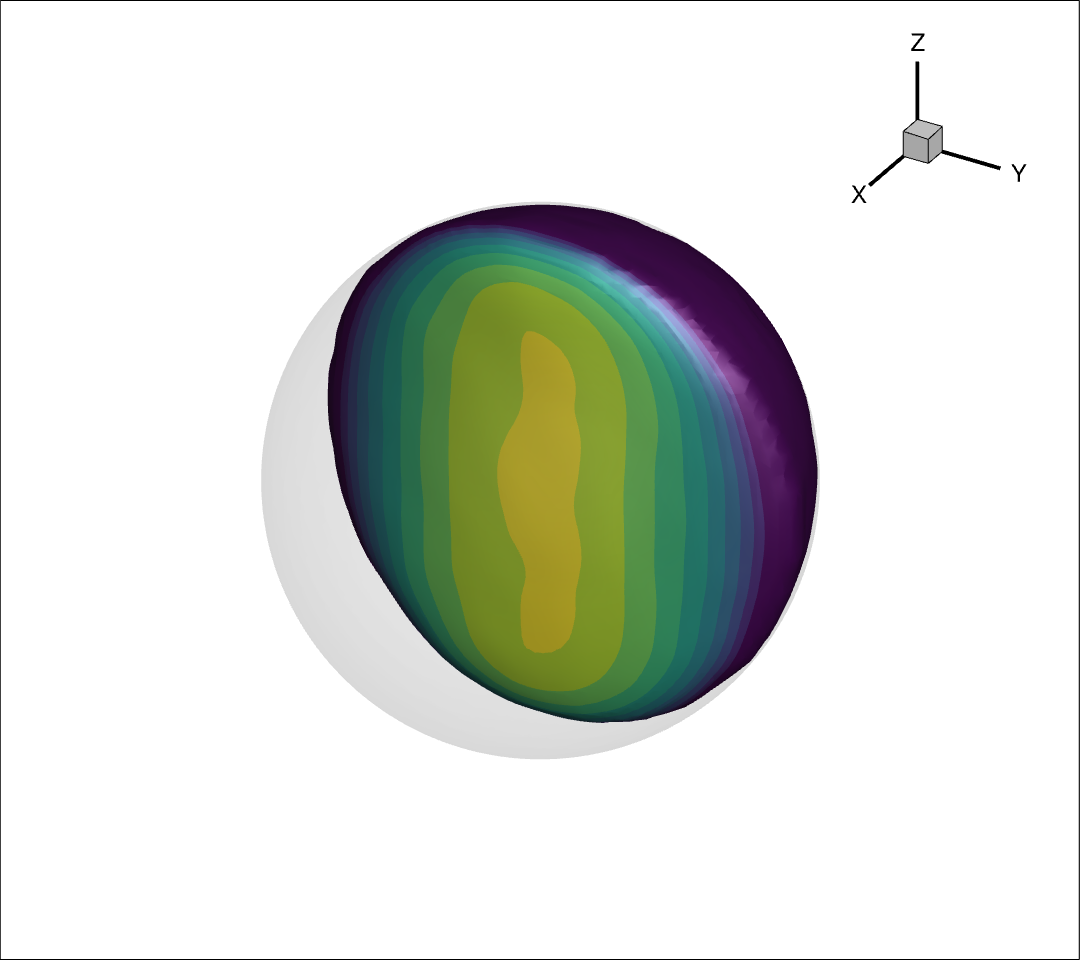}
        \includegraphics[width=0.24\textwidth, trim=15 15 15 15, clip]{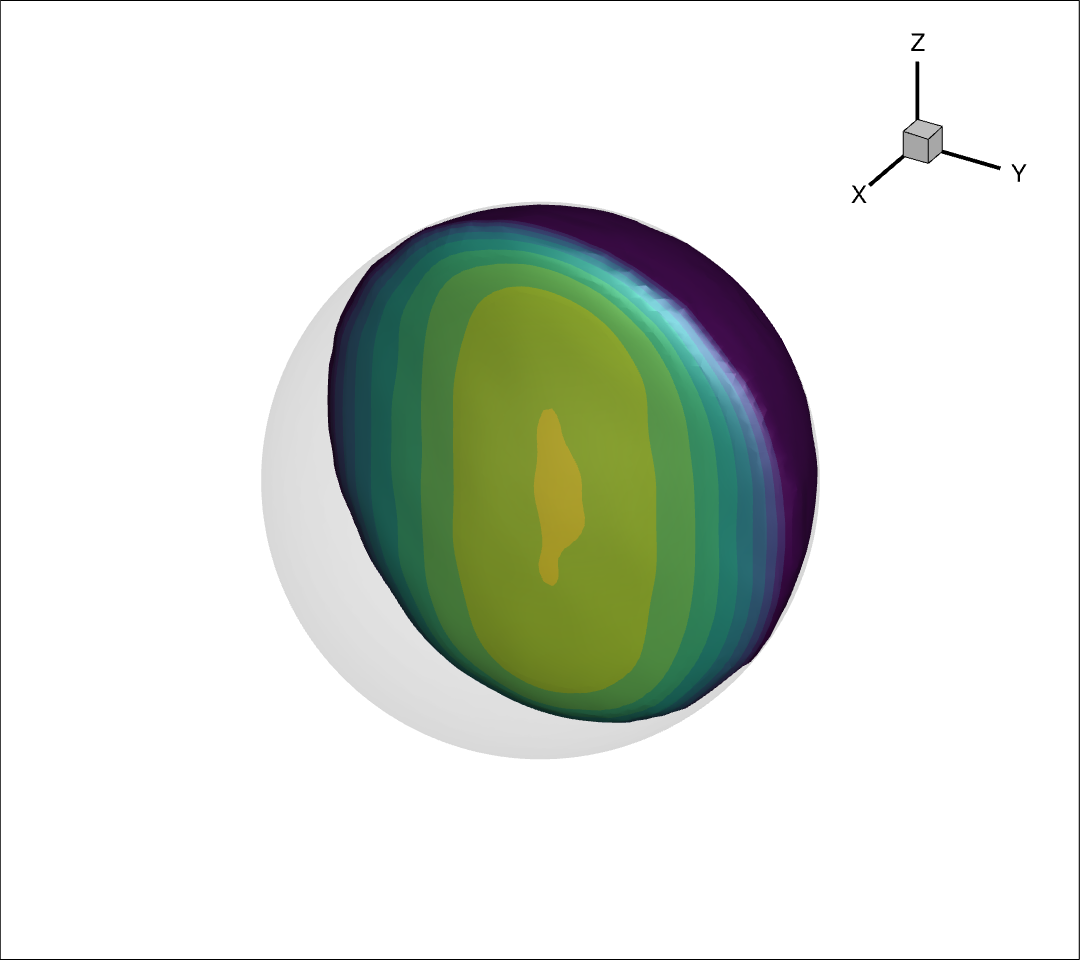}

        \vspace{-1em}
        \includegraphics[width=0.6\textwidth, trim=5 0 5 5, clip]{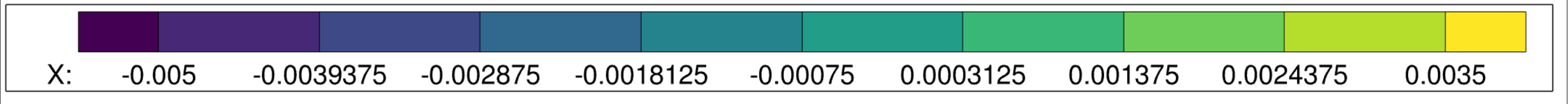}
        \vspace{-0.5em}

        \subcaptionbox{$t=69.70s$\label{fig:sub1}}[0.24\textwidth]{}
        \subcaptionbox{$t=71.21s$\label{fig:sub2}}[0.24\textwidth]{}
        \subcaptionbox{$t=72.73s$\label{fig:sub3}}[0.24\textwidth]{}
        \subcaptionbox{$t=74.24s$\label{fig:sub4}}[0.24\textwidth]{}
    \end{minipage}

    \caption{Snapshots of the gas-liquid interface oscillations from DIVA simulation. Gas bubble colored for $X$ levels.}
    \label{Fig:snapshot-interface}
\end{figure}

The EMM successfully captures the dominant sloshing forces and their oscillations, which are closely linked to the motion of the liquid's center of gravity (CoG), indicating that the model effectively represents the bulk fluid motion during large-amplitude sloshing in microgravity. However, CFD simulations reveal residual low-damping oscillations not reproduced by the EMM. These arise from interfacial dynamics at the fluid surface, as shown in the snapshots in Fig.~\ref{Fig:snapshot-interface}.

 We attribute the slow damping observed in the CFD results to the idealised wetting assumption: the use of a zero contact angle between the liquid and the tank wall. This simplification neglects critical energy dissipation mechanisms at the three-phase contact line. In reality, sloshing causes the contact line to move, inducing energy loss through contact line friction and contact angle hysteresis. The latter, defined as the difference between advancing ($\theta_A$) and receding ($\theta_R$) contact angles ($\theta_A > \theta_B$), is a well documented dissipative dynamic process \cite{huh1971hydrodynamic, dussan1979spreading, blake1993kinetics, quere2008wetting, snoeijer2013moving}, and its influence becomes particularly significant in low gravity environments, where surface tension dominates fluid behavior \cite{maeda2009slosh}. 
The EMM does not aim to capture these fine-scale interface effects, as it is fundamentally designed to model the motion of the sloshing mass CoG rather than detailed surface dynamics. Therefore, the absence of these low-damped oscillations in the model output is consistent with its intended level of abstraction. Based on its accurate prediction of the main sloshing forces and torques, we consider the parameter tuning successful and the model valid for representing the dominant dynamics of low-g sloshing.
Furthermore, the EMM offers a significant computational advantage. While CFD simulations require several hours to simulate a few hundred seconds of physical time, the EMM produces comparable force predictions within seconds. Specifically, simulating $150$ seconds of physical time with DIVA required $34.3$ CPU hours on 16 cores of an Intel\textregistered{} Xeon\textregistered{} Gold 6126 processor using a $64^3$ mesh. While the EMM completed the same simulation in $3.07 s$ on a single core of an Intel\textregistered{} Core\texttrademark{} i7-11850H processor.

\subsection{Closed-Loop with feedback control}\label{sec:closed-loop-results}

After tuning and validating the model under open-loop conditions, the next step is to evaluate its reliability in a closed-loop scenario. To remain within the model range of applicability, a representative satellite spin-up manoeuvre is selected as the test case. The satellite begins at rest and accelerates to a nominal angular velocity $\omega_{z,nom}=1.5rad/s$ around its spin axis over a time interval $t_{acc}=10s$, with a constant angular acceleration $\dot \omega_{z,nom}=0.15rad/s^2$.

\begin{figure}[!h]
\centering
\includegraphics[width=0.75 \textwidth]{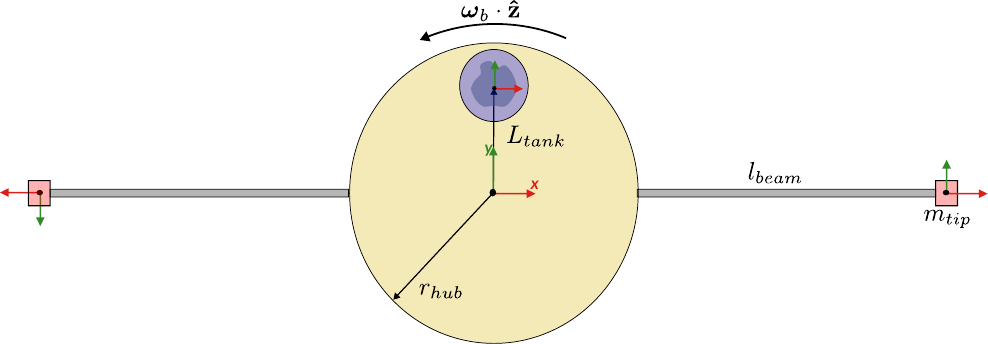}
\caption{Top view representation of the considered spinning satellite geometry.}
\label{Fig:sketch-satellite}
\end{figure}

The spacecraft features a cylindrical central hub with radius $r_{hub}$ and height $h_{hub}$. Two lateral rods of length $l_{beam}$, each terminating in a tip mass $m_{tip}$, extend from both ends of the hub. The spherical propellant tank, filled to $50\%$ capacity with 3M NOVEC 2704 as in the previous case, is positioned at a distance $L_{tank}$ from the hub center, oriented in the binormal direction relative to both the beam axis and the hub vertical axis. This configuration is illustrated in Fig. \ref{Fig:sketch-satellite}, while Table \ref{Tab:satellite-parameters} details the structural parameters. 

\begin{table}[h!]
\centering
\caption{Structural parameters of the spacecraft used in the closed-loop simulation.}
\begin{tabular}{llcl}
\toprule
\textbf{Component} & \textbf{Symbol} & \textbf{Description} & \textbf{Value} \\
\midrule
\multirow{4}{*}{Central hub $\mathcal{B}$}
  & $m_{hub}$ & Mass of central hub & $10\,\mathrm{kg}$ \\
  & $r_{hub}$ & Radius of hub & $0.4\,\mathrm{m}$ \\
  & $h_{hub}$ & Height of hub & $0.2\,\mathrm{m}$ \\
  & $J_{total}$ & Inertia tensor at CoM (in $\mathcal{F}_b$) &
    $\left[\begin{array}{ccc}
    0.5002 & 0 & 0 \\
           & 1.2404 & 0 \\
           &        & 1.6727
    \end{array}\right]\,\mathrm{kg\,m^2}$ \\
\midrule
\multirow{2}{*}{Lateral rods $\mathcal{A}$}
  & $l_{beam}$ & Beam lenght & $0.2\,\mathrm{m}$ \\
  & $m_{tip}$ & Tip mass & $1\,\mathrm{kg}$ \\
\midrule
Propellant tank $\mathcal{T}$ 
  & $L_{tank}$ & Position vector from hub center (in $\mathcal{F}_b$) & $[0,\,0.2667,\,0]\,\mathrm{m}$ \\
\bottomrule
\end{tabular}
\label{Tab:satellite-parameters}
\end{table}

\begin{figure}[!h]
    \centering
    \begin{subfigure}[b]{0.75\textwidth}
        \centering
        \includegraphics[width=\textwidth]{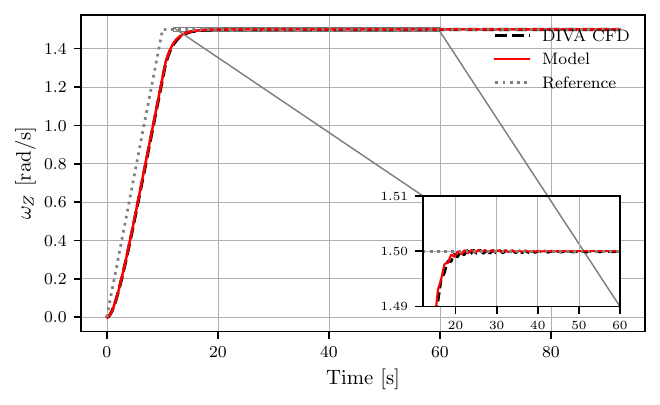}
        \caption{Controlled angular velocity from CFD and model.}
        \label{fig:sub1}
    \end{subfigure}
    \begin{subfigure}[b]{0.75\textwidth}
        \centering
        \includegraphics[width=\textwidth]{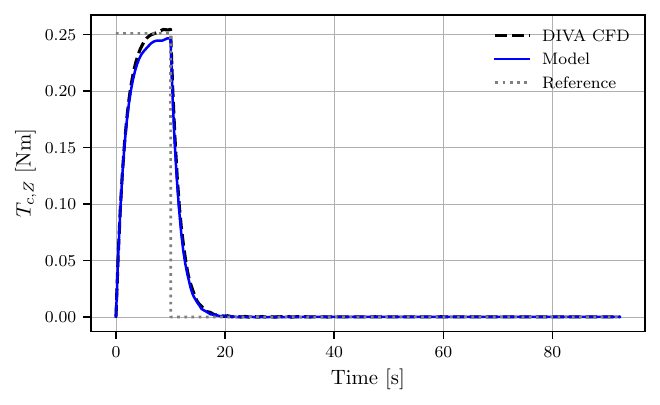}
        \caption{Control torque $T_{c,Z}$ from CFD and model.}
        \label{fig:sub3}
    \end{subfigure}
    \caption{Comparison of the controlled angular velocity and control torque from CFD and sloshing model during the closed-loop manoeuvre.}
    \label{Fig:closed-loop-control}
\end{figure}

\begin{figure}[!h]
    \centering
    \begin{subfigure}[b]{0.75\textwidth}
        \centering
        \includegraphics[width=\textwidth]{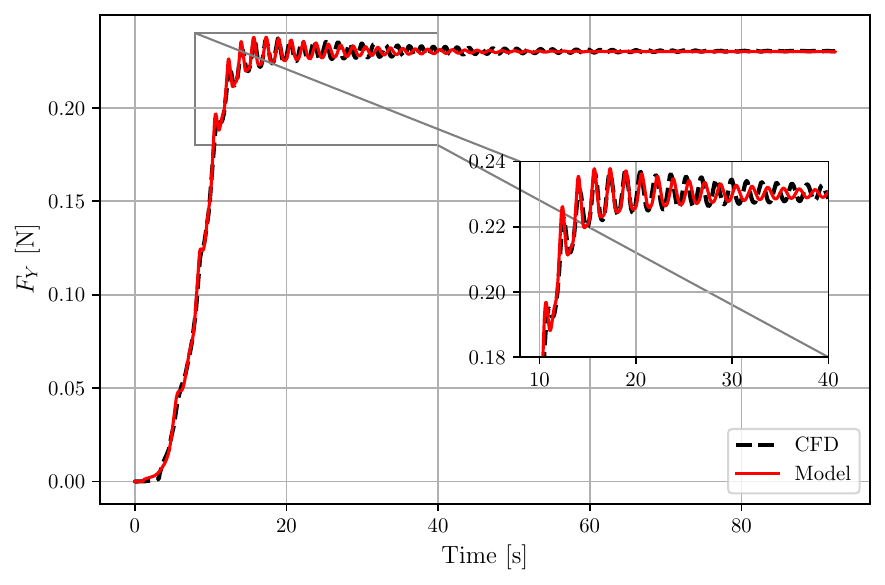}
        \caption{Force component $F_Y$ from CFD and model.}
        \label{fig:sub1}
    \end{subfigure}
    \begin{subfigure}[b]{0.75\textwidth}
        \centering
        \includegraphics[width=\textwidth]{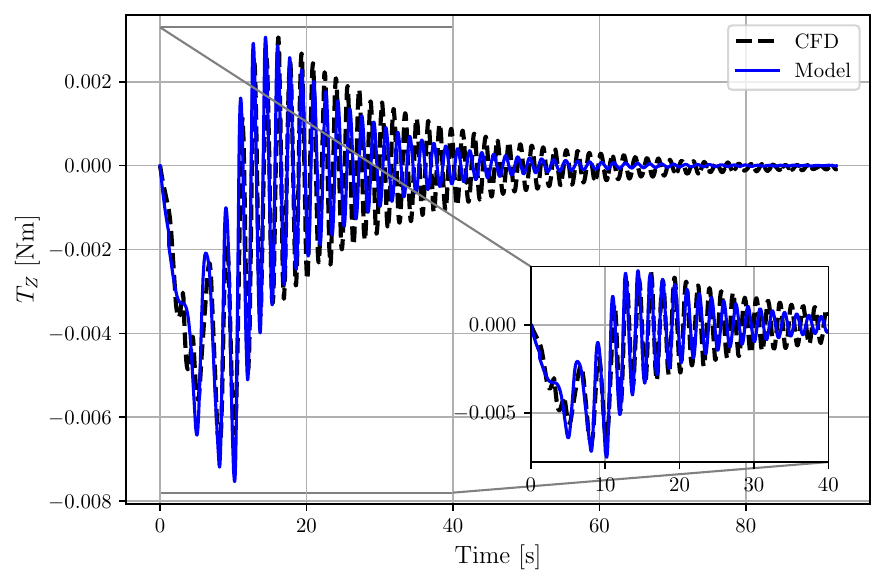}
        \caption{Torque component $T_Z$ from CFD and model.}
        \label{fig:sub3}
    \end{subfigure}
    \caption{Comparison of the main force and torque components from CFD and sloshing model during the closed-loop manoeuvre.}
    \label{Fig:closed-loop-forces}
\end{figure}

\noindent The maneuver results in Bond numbers $Bo_i=10.4$ and $Bo_c=155$, indicating stronger inertial effects compared to the open-loop case.\\
To isolate the fluid-structure interaction effects, the satellite structure is assumed rigid. Attitude control is implemented using a simple static proportional control law, designed to follow the prescribed reference angular velocity. The control input $u$ defines the torque applied around the spin axis and is defined as: 

\begin{equation}
u = \mathcal{K} \, \Delta \omega, \qquad \text{with} \quad
\left\{
\begin{aligned}
\Delta \omega &= \omega_{z,\text{nom}} - \omega_{z,\text{sim}}, \\
\mathcal{K} &= 2 \varepsilon \omega_n J_z,
\end{aligned}
\right.
\end{equation}

\noindent where $\omega_{z,\text{sim}}$ is the actual angular velocity of the satellite about the $z$-axis, $\omega_{z,\text{nom}}$ is the target spin rate, $J_z$ is the moment of inertia about the spin axis, and the control parameters are $\varepsilon=0.7$ and $\omega_n=0.06 rad/s$. Simulations are conducted using the procedure outlined in Fig. \ref{Fig:simulation-algorithm}, comparing results from the DIVA code and the EMM.
Figure \ref{Fig:closed-loop-control} presents the time evolution of the controlled angular velocity and the associated control torques. Figure \ref{Fig:closed-loop-forces} shows the sloshing-induced centrifugal forces $F_Y$ and disturbance torques about the spin axis $\hat{z}$. 

 Despite the higher Bond numbers and the modified acceleration profile induced by the controller, the EMM maintains good agreement with DIVA results. The model accurately captures the dominant sloshing forces and oscillations during the initial acceleration phase. As the system approaches steady-state, oscillations in the model damp more rapidly than in the CFD results, ultimately converging to the expected asymptotic behaviour, consistent with observations from the open-loop case. 
Notably, in this closed-loop configuration, the control action effectively suppresses free-surface oscillations. The liquid interface reaches a nearly quiescent state within approximately 80 seconds, highlighting the damping effect introduced by feedback control on sloshing dynamics.
These results confirm that, once tuned on open-loop data, the model remains reliable even under feedback-controlled conditions for this class of manoeuvres. This is particularly valuable given the substantial computational efficiency it offers: the model runs at least an order of magnitude faster than real time, supporting computationally intensive tasks such as MCMC analysis or high-iteration early-stage design studies. In this case, for $92s$ of simulated physical time, DIVA took $20h$ on $64$ cores for a total of 1,280 CPU hours on the Intel\textsuperscript{\textregistered} Xeon\textsuperscript{\textregistered} Gold 6126 processor using a $128^3$ mesh. For the EMM, it took $17.76 s$ on a single core of the Intel\textsuperscript{\textregistered} Core\texttrademark{} i7-11850H processor.

\section{Discussion and future works}

This study investigated the reliability of a reduced-order model for predicting liquid sloshing in low-gravity environments under closed-loop control of a satellite-propellant system. The Equivalent Mechanical Model (EMM), which represents the sloshing fluid as a particle constrained on a revolute surface, was selected for its ability to capture the bulk reorientation of liquid during large-amplitude sloshing in low gravity \cite{Zhou2015, Elke2024}.

The model was first calibrated using high-fidelity CFD data from the DIVA solver \cite{Dalmon2019}, then tested in a closed-loop simulation where it was coupled with the rigid satellite structure and a feedback control law. Results showed strong agreement with CFD predictions, reproducing the main sloshing forces and oscillations even under acceleration profiles different from those used in the tuning phase, despite the model's simplicity.

A key contribution of this work lies in the closed-loop validation of the EMM. While such models are often validated in open-loop, their behavior under feedback control can differ significantly. Verifying model performance in this context is essential for reliable control design and for ensuring simulations remain within the model's domain of applicability, avoiding misleading outcomes in more complex scenarios. In addition, the EMM offers a substantial computational advantage, with simulations running at least an order of magnitude faster than real time. This makes the model well suited for tasks involving many iterations, such as sensitivity analysis, uncertainty quantification, and early-stage design. 

Future work will incorporate structural flexibility to explore interactions between sloshing and flexible structural modes, which are particularly relevant for satellites with large appendages, where dynamic coupling may affect stability and control. In conclusion, the proposed control-structure-propellant framework provides a promising foundation for the integrated design and analysis of future space missions.

\section*{Acknowledgments}

This work was supported by the Open Space Innovation Platform (OSIP) of the European Space Agency (ESA), Contract No. 4000143096/23/NL/MGu/my; the von Karman Institute, Contract No. ARD2407; and the Chair SaCLaB2, resulting from the partnership between Airbus Defence and Space, Ariane Group, von Karman Institute and ISAE-SUPAERO, which co-funds the PhD thesis of Umberto Zucchelli. M.A. Mendez is funded by the European Research Council (ERC) under the European Union's Horizon Europe programme (RE-TWIST project, grant agreement No 101165479). The views expressed are those of the authors and do not necessarily reflect those of the European Union or the ERC.

%
%

\bibliographystyle{plain}

\end{document}